\documentclass[11pt,a4paper]{article}
\usepackage{jheppub}

\usepackage{graphicx}
\usepackage{setspace}

\begin{document}




\author{Olga Chekeres}



\affiliation{Department of Mathematics, University of Geneva,\\
 2-4 rue du Li\`evre, c.p. 64, 1211 Gen\`eve 4, Switzerland}

\emailAdd{Olga.Chekeres@unige.ch}

\title{ Quantum Wilson surfaces and topological interactions. 
}

\begin{abstract}   
{We introduce the description of a Wilson surface as a 2-dimensional topological quantum field theory with a 1-dimensional Hilbert space. On a closed surface, the Wilson surface theory defines a topological invariant of the principal $G$-bundle $P \to \Sigma$. Interestingly, it can interact topologically with 2-dimensional Yang-Mills and BF theories modifying their partition functions. We compute explicitly the partition function of the 2-dimensional Yang-Mills theory with a Wilson surface. The Wilson surface turns out to be nontrivial for the gauge group $G$ non-simply connected (and trivial for $G$ simply connected). In particular we study in detail the cases $G=SU(N)/\mathbb{Z}_m$, $G=Spin(4l)/(\mathbb{Z}_2\oplus\mathbb{Z}_2)$ and obtain a general formula for any compact connected Lie group. 
}




\end{abstract}

\maketitle


\section{Introduction}

The discussion of surface observables in gauge theories has been ongoing for quite some time. Wilson surfaces, domain walls, surface defects etc appear in many domains of physics and mathematics, from gauge theories to condensed matter. They have been studied extensively in literature \cite{CR02, CHWZ07, C02, G96, GK, K14, P14}. In most cases a 1-dimensional observable, namely a Wilson line \cite{W74, G81, W89, AFS88, BBS78, NR88, DP89, EMSS95, B11}, is generalized to 2 dimensions by introducing higher gauge fields defined on surfaces. Our approah is different, it is based as well on a definition of a Wilson line, but doesn\rq{}t involve introducing higher gauge fields. 

The basis for our construction is a 1-form standard gauge field taking values in a Lie algebra. 
A Wilson surface is defined by an orientable surface $\Sigma$ and a representation $R_\lambda$ of the gauge group $G$. In \cite{ACM15} we obtained its description as a 2-dimensional topological $\sigma$-model:
\begin{equation}\label{sigma}
S_\lambda(a,b,A)=  \int_\Sigma \, {\rm Tr}(b (d(A+a) + (A+a)^2))=\int_\Sigma \, {\rm Tr}(b F_{A+a}),
\end{equation}
where $\lambda\in\Lambda^*$ is the highest weight of the representation $R_\lambda$, $b$ is a scalar field taking values in $\mathfrak{g}^*$ and constrained to be a conjugate of $\lambda\in\Lambda^*$, $A$ is a background gauge field, and $a$ is an auxiliary gauge field. Interpreting $A+a$ as a new gauge field allows us to interpret a Wilson surface as an independent 2-dimensional BF-theory with a constraint on the B-field. 

In this article we start with the action functional \eqref{sigma} and canonically quantize it. Our first result is the partition function formula for this Wilson surface theory. To describe the formula we first recall some topological facts. Principal $G$-bundles $P\to \Sigma$ over a closed surface $\Sigma$ are classified by the elements $\gamma\in\pi_1(G)$ in the fundamental group of the gauge group $G$ \cite{S51, H94}. The gauge group can be represented as $G=\tilde{G}/\Gamma$, where $\tilde{G}$ is the universal cover of $G$ and $\Gamma \subset Z(\tilde{G})$ is a proper subgroup of the center of $\tilde{G}$. Since $\Gamma \cong \pi_1(G)$, for every element $\gamma\in\pi_1(G)$ in the fundamental group there exists a corresponding element $C_\gamma\in\Gamma\subset Z(\tilde{G})$ in the center of the covering group. Then the partition function for the particular equivalence class of principal bundles $P\to \Sigma$, defined by $\gamma\in \pi_1(G)$, is given by:
\begin{equation}\label{phase}
Z^{\Sigma}_{WS}(C_\gamma, \lambda) = \frac{\chi_{\lambda}(C_\gamma)}{d_\lambda}=e^{i\varphi_\gamma} \in U(1), 
\end{equation}
where $\chi_{\lambda}(C_\gamma)$ is a value of the character $\chi_\lambda$ on the element $C_\gamma$ and ${d_\lambda}$ is the dimension of the representation. This is a 2-dimentional topological quantum field theory with a 1-dimensional Hilbert space.  

Our second result is the description of topological interactions of Wilson surfaces with 2-dimensional topological gauge theories, namely with BF and Yang-Mills theories. Their partition functions on a surface $\Sigma$ are obtained by summation over all the classes of principal $G$-bundles defined over the given surface \cite{W91,W92, CMR94}.
 When we insert a Wilson surface into 2-dimensional Yang-Mills or BF, it interacts topologically with the background gauge theory, as the gauge connections $A$ and $A+a$ are defined on the same principal $G$-bundle $P\to\Sigma$. The presence of a Wilson surface modifies the partition function of the background theory multiplying by a phase \eqref{phase} the individual contributions for each class of principal bundles:
 \begin{equation}\label{phase1}
 Z^{interact}=\sum_{\gamma\in\pi_1(G)} Z^{backgr}(C_\gamma) \cdot e^{i\varphi_\gamma}.
 \end{equation}
 
 Next, we study concrete examples of 2-dimensional Yang-Mills theory with a Wilson surface for the gauge groups $G=SU(N)/\mathbb{Z}_m$ ($m$ divides $N$) and $G=Spin(4N)/(\mathbb{Z}_2\oplus\mathbb{Z}_2)$.\footnote{The examples of $G=U(1), SU(2), SO(3)$ were computed in \cite{ACM15}.} Since Yang-Mills in 2 dimensions is exactly solvable, we can obtain explicit formulas for the partition function in the presence of a Wilson surface. 
 
 In case of $G=SU(N)/\mathbb{Z}_m$ the fundamental group is $\pi_1(G)\cong \mathbb{Z}_m$ and the Wilson surface phase $e^{i\varphi_k}$ is defined by the angle:
 $$ 
 \varphi_k = \frac{2\pi k}{m}\cdot[\lambda],
 $$
where $k=0,1,...,m-1$ labels the elements of $\pi_1(G)$, and $[\lambda]\in\mathbb{Z}_m$ is an integer mod $m$ denoting the equivalence class of the highest weight $\lambda$ characterising the Wilson surface.
For $G=Spin(4N)/(\mathbb{Z}_2\oplus\mathbb{Z}_2)$ the fundamental group is $\pi_1(G)\cong \mathbb{Z}_2\oplus\mathbb{Z}_2$ and the Wilson surface is defined by the angle:
$$ 
 \varphi_{k_1,k_2} = \pi (k_1[\lambda_1]+k_2[\lambda_2]),
$$
where a pair $(k_1,k_2)$ labels the elements of $\pi_1(G)$, with $k_1,k_2 \in \{0,1\}$, and $[\lambda_1], [\lambda_2] \in \mathbb{Z}_2$ are integers modulo $2$ given by two linear combinations of the components of same highest weight $\lambda$ characterizing the Wilson surface.

Eventually, we obtain the formula of the partition function for 2D-YM with a Wilson surface for any compact connected Lie group $G=\tilde{G}/\Gamma$. In this case the Wilson surface phase is defined by the angle:
$$
\varphi_{k_1,...,k_i} = \sum_i k_i <\lambda, c_i>= 2\pi \sum_i \frac{k_i}{m_i}[\lambda_i].
$$
Here we account for the most general case, when the fundamental group of $G$  is given by a product of $i$ cyclic groups: $\pi_1(G)=\mathbb{Z}_{m_1}\times ...... \times \mathbb{Z}_{m_i}$. Then $m_i$ is the number of elements in $\mathbb{Z}_{m_i}$, the index $k_i=0,...,m_i-1$ labels the elements in the $i$-th factor, $k_ic_i\in\mathfrak{h}\subset Lie(\tilde{G})$ is an element of the Cartan subalgebra such that $e^{i\sum_i k_i c_i} = C_{k_1,...,k_i}\in \Gamma \cong \pi_1(G)$ is a central element of the covering group $\tilde{G}$, $\lambda\in \mathfrak{h}^*$ is the highest weight of the representation of $G$ characterizing the Wilson surface, $<,>$ is the invariant scalar product defined on $Lie(\tilde{G})$ and $[\lambda_i]\in \mathbb{Z}_{m_i}$ are integers modulo $m_i$ given by $i$ linear combinations of the components of same highest weight $\lambda$.

For a closed surface the Wilson surface is nontrivial for $G$ non-simply connected, and it is not visible ($e^{i\varphi}=1$) for $G$ simply connected. Also the value of $\lambda$ plays a role: for $\lambda$ being the highest weight of a representation of the gauge group $G$ itself the Wilson surface is trivial, and it is nontrivial if $\lambda$ labels a representation of the universal cover $\tilde{G}$ which does not descend to $G$. 
On a closed surface, the partition function of the Wilson surface is a topological invariant of the principal $G$-bundle.

\vskip 0.2cm

{\bf Acknowledgements.} Our deepest gratitude is to A. Alekseev for inspiration throughout this work. We also thank D. Nedanovski, who participated in the early stages of the project and independently confirmed our computation for $G=SU(N)/\mathbb{Z}_N$, F. Valach for illuminating discussions and all the inhabitants of Villa Battelle Math Department in Geneva for inspiring atmosphere. 
Our research was supported in part by the grant 178794, the grant MODFLAT of the European Research Council (ERC) and the NCCR SwissMAP of the Swiss National Science Foundation. 


\section{Wilson surface theory}

\subsection{Wilson surface observables}


Recall the construction of Wilson surface observables from \cite{ACM15}. Let $G$ be the gauge group, $\mathfrak{g}$ its Lie algebra, $(x,y) \to {\rm Tr}(xy)$ an invariant scalar product on $\mathfrak{g}$ and $P$ a principal $G$-bundle over a surface $\Sigma$. We denote by $\mathfrak{h} \subset \mathfrak{g}$ a Cartan subalgebra  and by $\Lambda^* \subset \mathfrak{h}^*$ the weight lattice. 

A Wilson surface observable is described by an auxiliary 2-dimensional gauge theory on the surface $\Sigma$. The fields in this theory are a $\mathfrak{g}^*$-valued scalar field $b$ and a $\mathfrak{g}$-valued 1-form $a$. The action depends on the following data: the background gauge filed $A$ and the weight $\lambda \in \Lambda^*$. For a trivial $G$-bundle it is given by
%
\begin{equation}\label{PSM}
\begin{array}{lll}

S_\lambda(a,b,A)&= &\int_\Sigma \, {\rm Tr} (b(F_A -(dgg^{-1}+A)^2 + (dgg^{-1}+A+a)^2) \\
&=& \int_\Sigma \, {\rm Tr}(b (d(A+a) + (A+a)^2)),
\end{array}
\end{equation}
where we identifed $\mathfrak{g}^*$ with $\mathfrak{g}$ using the scalar product and integrated by parts using the equality ${\rm Tr} b [dgg^{-1}, a] = {\rm Tr} [b, dgg^{-1}] a = - {\rm Tr} (db) a$.
The field $b=g\lambda g^{-1}$ belongs to the same conjugacy class as the fixed element $\lambda$, the combination $A+a$ is a new gauge field.
Note that integrating out $a$ in \eqref{PSM} yields the Diakonov-Petrov action \cite {DP95, ACM15} for a Wilson surface: 
\begin{equation}
S_{DP} =\int_\Sigma \, {\rm Tr} (b(F_A -(dgg^{-1}+A)^2).
\end{equation}

The construction \eqref{PSM} also works on nontrivial bundles. $A\in \Omega^1(P, \mathfrak{g})$ is the connection on $P$, its curvature $F_A \in \Omega^2_{hor}(P, \mathfrak{g})^G$ is a horizontal 2-form taking values in $\mathfrak{g}$. The auxiliary gauge field $a$ is such that $a \in \Omega^1_{hor}(P, \mathfrak{g})^G$ and the sum $A+a$ defines a new connection on $P$ with a curvature $F_{A+a} = d(A+a) + (A+a)^2$, $F_{A+a} \in \Omega^2_{hor}(P, \mathfrak{g})^G$. The field $b$ takes values in $\Omega^0_{hor}(P, \mathfrak{g})^G$. The combination ${\rm Tr}(bF_{A+a})$ is then a basic $2$-form which decends to $\Sigma$. One can show this in the following way. The $2$-form $bF_{A+a}$ is $G$-equivariant, i.e. with respect to a gauge transformations by $h \, : \, \Sigma \to G$:
$$ g\mapsto hg \, , \, A \mapsto hAh^{-1} - dhh^{-1} \, , b\mapsto hbh^{-1} \, , \, a \mapsto hah^{-1},$$
it transforms as $b^hF_{A^h+a^h}=hbF_{A+a}h^{-1}$, yeilding ${\rm Tr}(b^hF_{A^h+a^h})={\rm Tr}(bF_{A+a})$.
Acting by the contraction we obtain:
$$\imath_{\xi^{\sharp}} {\rm Tr}(bF_{A+a}) = \imath_{\xi^{\sharp}}{\rm Tr} \, (b (F_A +a^2 +dgg^{-1}a + adgg^{-1} +Aa +aA)) = {\rm Tr} \, (b (- \xi a + a\xi  +\xi a - a\xi)) = 0,
$$
where $\xi\in \mathfrak{g}$ induces the fundamental vector field $\xi^{\sharp} \in \mathfrak{X}(\Sigma)$, and we have used that $\imath_{\xi^{\sharp}}(dg) = -\xi g$, $\imath_{\xi^{\sharp}}A = \xi$ (by definition of connection), $\imath_{\xi^{\sharp}}F_A = 0$, $\imath_{\xi^{\sharp}}a =0$ ($F_A$ and $a$ are horizontal). This computation proves that $G$-invariant form ${\rm Tr}(bF_{A+a})$ is horizontal, and hence basic. 

The meaning of $\lambda \in \Lambda_+^*$ is as follows. 
The integral weights of the representations of $G$ form the weight lattice $\Lambda^* \subset \mathfrak{h}^*$. Dominant integral weights $\Lambda_+^*\subset \Lambda^*$ are in one to one correspondence with irreducible representations of $G$ \cite{Z78, B79}. The element $\lambda \in \Lambda_+^*$ is the highest weight of some representation of $G$, it is a parameter characterising the Wilson surface. For example, for $G=SU(2)$ or $G=SO(3)$ we talk about a Wilson surface of spin $\lambda$. 
Note that in case when $G$ is not simply connected but is a quotient $G=\tilde{G}/\Gamma$, where $\tilde{G}$ is its universal cover and $\Gamma\subset Z(\tilde{G})$ is a subgroup of the center of $\tilde{G}$, the weight lattices are related as $\Lambda^*_G \subset \Lambda^*_{\tilde{G}}$. A representation of $\tilde{G}$ can be considered as projective representation of $G$, and $\lambda$ is allowed to take values in $\Lambda^*_{\tilde{G}}$, and, as we will see later, these are exactly the values which describe the presence of nontrivial Wilson surfaces.

\subsection{Quantum Wilson surfaces}

Rewriting the action for a Wilson surface \eqref{PSM} with respect to the new connection $A+a$ gives us a 2-dimensional $BF$ theory \cite{BBRT91} described on $\Sigma$:

\begin{equation}\label{1BF}
S_\lambda(a,b,A)=\int_\Sigma \, {\rm Tr}(b F_{A+a}).
\end{equation}

 Recall canonical quantization of $BF$-theory on a surface \cite{W91,W92}. For simplicity let first $\Sigma$ be a cylinder $C$, the $G$-bundle $P$ will be necessarily trivial. We chose space and time coordinates $(x,t)$ in a way that the boundary of $C$ is given by two closed curves $\gamma_1$ and $\gamma_2$, situated on equal time slices, and $x$ is a periodic coordinate of period $L$. We associate to $\gamma_i$ a gauge invariant wave function $\psi(A)$ which is a function of holonomy of $A$ around $\gamma_i$:
 $$ \psi [U_i] = \psi [Pe^{\int_0^LdxA_1}].$$
 
 The Hilbert space $\mathcal{H}_{\gamma}$ of such a theory is given by $G$-invariant $L^2$ functions on $G$. $\mathcal{H}_{\gamma}$ admits a natural basis in terms of characters of representations, and any wave function $\psi(A) \in \mathcal{H}_{\gamma}$ has an expansion in characters $\chi_R(U)$, where $R$ is a representation. The boundaries are oriented, the wave functions on the incoming and outgoing boundary components are denoted by $\overline{\chi_R}(U_i)$ and $\chi_R(U_j)$ respectively.

 In BF-theory Hamiltonian vanishes (as expected for a topological field theory), so the partition function reduces to:
 $$
 Z^C_{BF}(U_1, U_2) = \sum_R \overline{\chi_R}(U_1)\chi_R(U_2).
 $$
 
For a generic surface with genus $g$ and $r$ boundary components the BF partition function reads:
 $$
 Z^{\Sigma}_{BF}(U_1,...,U_r) = \sum_R d_R^{2-2g-r}\chi_R(U_1)...\chi_R(U_r),
$$
where $d_R$ is the dimension of the representation and all the boundaries are chosen to be outgoing.  

To obtain the formula for a closed surface we proceed as follows. 
The partition function will necessarily depend on the equivalence class of $G$-bundle over $\Sigma$. 
Recall the classification of principal $G$-bundles over $\Sigma$ by the elements of the fundamental group of $G$: $\pi_1(G) \cong \Gamma\subset Z(\tilde{G})$. 
Consider a surface with just one puncture, i.e. one boundary component. Gluing this puncture to an infinitesimal disc yeilds a closed surface, and this operation is described by identifying $U=C_i$, where $C_i\in \Gamma$ is a central element of $\tilde{G}$.  

The contribution to the partition function of each class $[P]$ of a principal $G$-bundle over the surface is given by:
 \begin{equation}\label{pf_P}
 Z^{\Sigma}_{BF}(C_i) = \sum_R d_R^{1-2g}\chi_R(C_i).
 \end{equation}
 
 The total partition function for BF-theory on a closed surface $\Sigma$ is then a sum over equivalence classes of principal $G$-bundles over $\Sigma$:
 \begin{equation}\label{pf}
 Z^{\Sigma}_{BF} = \frac{1}{\#\Gamma}\sum_{C_i\in \Gamma} \sum_R d_R^{1-2g}\chi_R(C_i),
 \end{equation}
 where $\#\Gamma$ is the cardinality of $\Gamma$.
 
 Note that the sum over $R$ in \eqref{pf} converges only for surfaces $\Sigma$ with genus $g>1$. 
 
Now we explain how to construct the partition function for a Wilson surface.  In contrast with BF-theory, the $\mathfrak{g}^*$-valued field $b=g\lambda g^{-1}$ is now the conjugation of the same fixed element $\lambda \in \Lambda^*_{\tilde{G}}$. The Hilbert space becomes one-dimensional choosing one representation $R_{\lambda}$, and the partition function is just a phase. The normalization of the states is such that $||<\chi_\lambda(U)|\chi_\lambda(U)>||^2=1$. 
 
 The state corresponding to a disc is given by
 \begin{equation}
 Z^{disc}_{WS}(U, \lambda) = \chi_{\lambda}(U). 
\end{equation}
The orientation of the disc chosen in a way that $U$ is a holonomy of the connection $A+a$ around an outgoing boundary.
 
The Wilson surface on a pair of pants has the formula:
\begin{equation}
 Z^{p-o-p}_{WS}(U_1, U_2, U_3, \lambda) = \overline{\chi_{\lambda}}(U_1)\chi_{\lambda}(U_2)\chi_{\lambda}(U_3), 
\end{equation}
where $U_1$, $U_2$, $U_3$ are holonomies of $A+a$ around one incoming and 2 outoing boundary components.

Any other orientable surface can be obtained by gluing those elementary components together. 
The partition function for a surface of arbitrary genus with $r$ boundary components is a product of $r$ states living on the boundaries (here chosen to be outgoing):
$$
Z^{\Sigma}_{WS}(U_1,....,U_r, \lambda) =  \chi_{\lambda}(U_1)......\chi_{\lambda}(U_r). 
$$

The expression for a closed surface for a particular class $[P]$ of principal bundles $P\to \Sigma$ is 

\begin{equation}\label{WS_P}
Z^{\Sigma}_{WS}(C_i, \lambda) = \frac{\chi_{\lambda}(C_i)}{d_\lambda}. 
\end{equation}

For a nontrivial central element $C_i$ this is an element of $U(1)$: $Z^{\Sigma}_{WS}(C_i, \lambda)\in U(1)$. In case when $P$ is a trivial bundle, the Wilson surface is always trivial: $Z^{\Sigma}_{WS}(P_{triv}, \lambda)=1$.


\section{Topological interactions with 2-dimensional gauges theories}

In general case our observable could be understood as a surface defect embedded into a higher dimensional space-time. But in this paper we want to test it in the context of 2-dimensional gauge theories, which can be solved exactly. In this case the Wilson surface is a \lq\lq{}global\rq\rq{} observable, defined on the entire 2-dimensional space-time $\Sigma$.

\subsection{BF theory with a Wilson surface}

The action functional for BF theory with a Wilson surface is:

\begin{equation}\label{bf_ws}
S^{\lambda}_{BF}(A, a, B, b) = \int_\Sigma \, {\rm Tr}(B F_A)+\int_\Sigma \, {\rm Tr}(b F_{A+a}),
\end{equation}
where $B\in \Omega^0(P, \mathfrak{g}^*)^G$, $A\in \Omega^1(P, \mathfrak{g})$ is the background gauge field, $F_A \in \Omega^2_{hor}(P, \mathfrak{g})^G$ its curvature. 

These two BF theories are not completely independent, they interact topologically: the connections $A$ and $A+a$ are defined on the same principal $G$-bundle, so the characters $\chi_R(U(A))$ and $\chi_{\lambda}(U(A+a))$ are taken on the same central element. 

Then the partition function for a closed surface with a Wilson surface of weight $\lambda$ is obtained by taking a product of partition functions defined in the previous section for each class $[P]$ and then summing over all the equivalence classes:
\begin{equation}\label{pf_bf_ws}
 Z_{BF}^{\lambda}= \frac{1}{\#\Gamma}\sum_{C_i\in \Gamma} \frac{\chi_{\lambda}(C_i)}{d_\lambda} \sum_R d_R^{1-2g}\chi_R(C_i).
 \end{equation}

\subsection{2D-YM theory with a Wilson surface}

Consider 2D-YM in the first order formalism. The action functional for the theory with a Wilson surface is:

\begin{equation}\label{ym}
S^{\lambda}_{YM}(A, a, B, b) = \int_\Sigma \, {\rm Tr}( BF_A + \frac{e^2}{2} B^2 d^2\sigma)+\int_\Sigma \, {\rm Tr}(b F_{A+a}),
\end{equation}
where $B$ is an auxiliary field taking values in $\mathfrak{g}^*$ and $d^2\sigma$ is the area element on $\Sigma$. Again, we see that the action splits into two theories interacting topologically through the connections $A$ and $A+a$ defined on the same principal bundle. The partition function for each class $[P]$ will be a product of the partition function for 2D-YM and the partition function for the Wilson surface. 

The Hamiltonian of the first theory is $H=\frac{e^2}{2}{\rm Tr}B^2$. The Hamiltonian of the second theory vanishes. The basis for the Hilbert space is given by gauge invariant functions $\psi_R(A,a) = \chi_R(U(A))\chi_{\lambda}(U(A+a))$, where $R$ runs through the irreps of $G$, $\lambda$ choses one irrep of $G$, and $U(A)$, $U(A+a)$ are holonomies of the connections $A$ and $A+a$ respectively. The eigenvalues of the Hamiltonian on $\psi(A,a)$ are given by quadratic Casimir $C_2(R)$ of the representation $R$, just like for the 2D YM without Wilson surface, as only the Hamiltonian of 2D YM contributes to the total theory. 

Then the time evolution operator takes value $e^{-\tau C_2(R)}$ on the functions $\psi_R(A,a)$, where $\tau = \frac{e^2}{2} \sigma$ absorbs the YM coupling constant $e^2$ and the area of the surface $\sigma$.

The partition function for a closed surface with a Wilson surface of weight $\lambda$ is given by the following formula:
\begin{equation}\label{pf_ym_ws}
 Z^{\lambda}_{YM}(\tau) = \frac{1}{\#\Gamma}\sum_{C_i\in \Gamma} \sum_R d_R^{1-2g}e^{-\tau C_2(R)}\chi_R(C_i)\frac{\chi_{\lambda}(C_i)}{d_\lambda}.
 \end{equation}
 
 \section{Exact results for 2D-YM theory interacting with a Wilson surface}
 
Yang-Mills theory in 2 dimensions is exactly solvable \cite{M75, B80, KK80, KK81, K81, GKS89, R90, F90, F91, BT92}, this allows us to obtain explicit formulas for partition function in the presence of a Wilson surface. In \cite{ACM15} we computed the partition functions for 2D Yang-Mills with a Wilson surface for the gauge groups $U(1)$, $SU(2)$ and $SO(3)$. Now we are going to generalize this result to $G$ being any compact connected Lie group. 
 
\subsection{SU(N) and the groups covered by SU(N)}

To visualize the result of topological interactions with a Wilson surface, we first perform a detailed computation for the case of $\tilde{G}=SU(N)$.  
The center of $SU(N)$ is given by:
$Z(SU(N)) = \{e^{\frac{2\pi i k}{N}}Id_N \, | \, k=0, ..., N-1\}=\mathbb{Z}_N$. And the subgroups of the centre are $\Gamma = \mathbb{Z}_m$ where $m$ devides $N$. 
We consider $G=SU(N)/\mathbb{Z}_m$. 

The rank of $SU(N)$ is equal to $N-1$, i.e. the basis of Cartan subalgebra $\mathfrak{h}$ has $N-1$ elements.  In the defining presentation the basis of $\mathfrak{h}$ is given by: $h_n=\rm{diag}(0.......,1,-1,......0)$ with matrix elements $h_{nn}=1$, $h_{n+1,n+1}=-1$. Then any element  $h\in\mathfrak{h}$ can be represented in terms of the basis as $h=\sum_{n=1}^{N-1} a_n h_n$, with $a_n\in\mathbb{R}$ linear coefficients. 
Exponentiating elements $h\in\mathfrak{h}$ we obtain the maximal torus of $SU(N)$: $H= e^{ih}= e^{i\sum a_n h_n}= \rm{diag}(e^{i\theta_1}, e^{i\theta_2},....,e^{i\theta_{N-1}}, e^{-i\sum_{i=1}^{N-1}\theta_i})\in T$.  

The center $Z(SU(N))$, and hence its proper subgroup $\Gamma \subset Z(SU(N))\subset T$, is a subgroup of the maximal torus: $\Gamma  \ni C_k=e^{ i \theta_k}Id_N\in T$, with $\theta_k=2\pi k/m$.

Consider the elements of the Cartan subalgebra $c_k = \rm{diag} (\theta_k, \theta_k,......,-(N-1)\theta_k)_{N\times N} \in \mathfrak{h}$, such that $C_k=e^{ic_k}\in Z(SU(N))$. In terms of the basis of $\mathfrak{h}$ they are given as follows:

$c_k = \rm{diag} (\theta_k, \theta_k,......,-(N-1)\theta_k) = \theta_k\cdot \rm{diag} (1, 1,.......,1,-(N-1))=\theta_k \cdot \sum_{n=1}^{N-1}nh_n$. Then the central elements of $SU(N)$ are given by $C_k=e^{i\theta_k \sum_{n=1}^{N-1}nh_n}\in Z(SU(N))$.

The irreducible representations of $SU(N)$ are labeled by highest weights with $N-1$ independent elements: $\mu = (\mu_1, ..., \mu_{N-1})$. 

The central elements in the representation $R_{\mu}$ of highest weight $\mu$ are obtained as:
$R_{\mu}(e^{i\theta_k \sum_{n=1}^{N-1}nh_n}) = e^{i\theta_k \sum_{n=1}^{N-1}nR_{\mu}(h_n)}$. 

The natural choice for the basis of $R_{\mu}$ is in terms of the weight vectors $v_i$ with $v_1$ the highest weight vector. In this basis $R_{\mu}(h_n)$ are diagonal and yield weights while acting on the basis vectors.
 A central element $C_k$ is a multiple of identity, therefore $R_{\mu}(C_k)$ has to be a multiple of identity as well, so it\rq{}s enough to compute it just on the highest weight vector:

\begin{equation}\label{R_SU(N)}
R_{\mu}(C_k) = e^{i\frac{2\pi k}{m}\sum_{n=1}^{N-1}n\mu_n}\cdot Id_{d_{R_\mu}}.
\end{equation}
The linear combination $\sum_{n=1}^{N-1}n\mu_n$ is an integer, but the expression \eqref{R_SU(N)} depends only on the value of this sum modulo $m$, as $e^{i\frac{2\pi k}{m}\sum_{n=1}^{N-1}n\mu_n}=e^{i\frac{2\pi k}{m}(\sum_{n=1}^{N-1}n\mu_n+m)}$. This allows us to define the equivalence classes of the highest weight $\mu$: 
\begin{equation}\label{class}
[\mu] \equiv [\sum_{n=1}^{N-1}n\mu_n] \in \mathbb{Z}_m.
\end{equation}

Note that the irreps of $\tilde{G}$ descend to the irreps of $G$ if $\sum_{n=1}^{N-1}n\mu_n=0$ mod $m$. 
In terms of weight lattices $\Lambda^*_G \subset \Lambda^*_{\tilde{G}} \subset \mathfrak{h}^*$, where $\Lambda^*_{\tilde{G}}$ is the weight lattice for $SU(N)$, $\Lambda^*_G$ is the weight lattice for $G=SU(N)/\mathbb{Z}_m$.

The characters of the central elements in the representations $R_\mu$ are as follows:

\begin{equation}
\chi_{R_{\mu}}(C_k) = Tr(e^{i\frac{2\pi k}{m}\sum_{n=1}^{N-1}n\mu_n}\cdot Id_{d_{R_\mu}}) = {d_{R_\mu}}\cdot (e^{i\frac{2\pi k}{m}})^{[\mu]}.
\end{equation}

We keep the notation $Z_{YM}(\tau)$ for the partition function of the free 2D-YM theory and $Z^{\lambda}_{YM} (\tau)$ for the theory with a Wilson surface of the highest weight $\lambda$.

Without Wilson surface the partition function for $SU(N)/\mathbb{Z}_m$ is given by 
\begin{equation}
 Z_{YM}(\tau) = \frac{1}{m}\sum_{k=0}^{m-1} \sum_{R_{\mu}(SU(N))} {d_{R_\mu}}^{1-2g}e^{-\tau C_2(R_{\mu})}\chi_{R_\mu}(C_k),
 \end{equation}
 where the sum is over the representations $R_\mu$ of $SU(N)$, and $k$ labels central elements in the subgroup $\Gamma$. 
 
 The computation gives the following result:
 \begin{equation}
 Z_{YM}(\tau) = \frac{1}{m}\sum_{k=0}^{m-1} \sum_{R_{\mu}(SU(N))} d_{R_\mu}^{2-2g}e^{-\tau C_2(R_{\mu})}(e^{i\frac{2\pi k}{m}})^{[\mu]} 
 = \sum_{R_{\mu}(SU(N))} d_{R_\mu}^{2-2g}e^{-\tau C_2(R_{\mu})}\frac{1}{m}\sum_{k=0}^{m-1} (e^{i\frac{2\pi k}{m}})^{[\mu]},
 \end{equation}
 where the sum over $k$ is equal to $m$ for $[\mu]=0$ and zero otherwise. The condition corresponds to those representations of $SU(N)$ in which the central elements are all trivial, that is to the representations of $SU(N)/\mathbb{Z}_m$:
 
\begin{equation}
 Z_{YM}(\tau)=\sum_{R_{\mu}(G=SU(N)/\mathbb{Z}_m)} d_{R_\mu}^{2-2g}e^{-\tau C_2(R_{\mu})}.
 \end{equation}
 
 Now let us introduce a Wilson surface of weight $\lambda$:  
\begin{equation}\label{+WS}
\begin{array}{lll}
 Z^{\lambda}_{YM}(\tau) &=& \frac{1}{\#\Gamma}\sum_{k=0}^{m-1} \sum_{R_\mu(SU(N))} d_{R_\mu}^{1-2g}e^{-\tau C_2(R_{\mu})}\chi_{R_\mu}(C_k)\frac{\chi_{\lambda}(C_k)}{d_\lambda} \\
 &=& \frac{1}{m}\sum_{k=0}^{m-1} \sum_{R_\mu(SU(N))} d_{R_\mu}^{1-2g}e^{-\tau C_2(R_{\mu})}\frac{d_\lambda\cdot (e^{i\frac{2\pi k}{m}})^{[\lambda]}}{d_\lambda}d_R\cdot (e^{i\frac{2\pi k}{m}})^{[\mu]}\\
 &=& \frac{1}{m}\sum_{k=0}^{m-1} \sum_{R_\mu(SU(N))} d_{R_\mu}^{2-2g}e^{-\tau C_2(R_{\mu})}(e^{i\frac{2\pi k}{m}})^{[\mu]+[\lambda]},
 \end{array}
\end{equation}
where $[\lambda] = [\sum_{n=1}^{N-1}n\lambda_n] \in \mathbb{Z}_m$ are equivalence classes of the Wilson surface weight $\lambda$. 
 In more detail, we consider a quotient map $\Lambda^*_{\tilde{G}}\ni \lambda \mapsto (\sum_{n=1}^{N-1}n\lambda_n)_{mod \, m} \in \mathbb{Z}_m $ and the highest weights for Wilson surfaces will belong to equivalence classes $[\lambda] \in \Lambda^*_{\tilde{G}}/\Lambda^*_G \cong \mathbb{Z}_m$.
Note that in case when $\lambda \in \Lambda^*_G$, i.e. $\sum_{n=1}^{N-1}n\lambda_n=0$ mod $m$, the Wilson surface is not visible:
$Z^{\lambda} (\tau)=Z(\tau)$. 

The sum over $k$ in \eqref{+WS} is different from zero only for $[\mu]+[\lambda]=0$, and the partition function formula for 2D-YM with a Wilson surface yields:
$$ Z^\lambda_{YM} (\tau) = \sum_{\mu\in\Lambda^*_{\tilde{G}} \,, \, [\mu + \lambda]=0} dim_{R_{\mu}}^{2-2g} e^{-\tau C_2(R_{\mu})}=\sum_{\mu \in [-\lambda]} dim_{R_{\mu}}^{2-2g} e^{-\tau C_2(R_{\mu})},$$
where the sum now goes over the representations $R_{\mu+\lambda}$ of $G=SU(N)/\mathbb{Z}_m$ of highest weights $\mu + \lambda$, i.e. $[\mu]=[-\lambda]\in \Lambda^*_{\tilde{G}}/\Lambda^*_G$.

Note that in case when $G=SU(N)$, i.e. the gauge group is simply connected, the presence of the Wilson surface makes no impact on the partition function. Let us look at this situation in more detail. 
There is just one class of principal $SU(N)$-bundles over a surface $\Sigma$ -- trivial $SU(N)$-bundle.  
The $SU(N)$ partition function without Wilson surface is given by:
\begin{equation}
 Z^{SU(N)}_{YM}(\tau) = \sum_{R_{\mu}} d_{R_\mu}^{1-2g}e^{-\tau C_2(R_{\mu})}\chi_R(e) = \sum_{R_{\mu}} d_{R_\mu}^{2-2g}e^{-\tau C_2(R_{\mu})},
 \end{equation}
 where $e$ is identity. 
 
 And adding a Wilson surface of weight $\lambda$ leaves the partition function unchanged:
 \begin{equation}
 Z^{\lambda, SU(N)}_{YM}(\tau) = \sum_{R_\mu} d_{R_\mu}^{1-2g}e^{-\tau C_2(R_{\mu})}\chi_R(e)\frac{\chi_{\lambda}(e)}{ d_{\lambda}}=  \sum_{R_\mu} d_{R_\mu}^{2-2g}e^{-\tau C_2(R_{\mu})}.
\end{equation}

\subsection{Generalization for any compact connected Lie group}
 
The result explained in the explicit example of the previous section remains valid for all compact connected Lie groups. All of them (with the exception of exceptional ones) have as a universal cover one of the following groups: $SU(N)$, $Spin(N)$, $Sp(N)$ and can be obtained by taking a quotient by a subgroup $\Gamma$ of the center. 

The case of $\tilde{G} = SU(N)$ has been discussed in the previous section. 
For $Spin(N)$, $N\ge3$ the data is as follows:
\begin{equation}
Z(Spin(N)) = \left\{ \begin{array}{ll}
\mathbb{Z}_2 & \textrm{if $N=2l+1$, $\Gamma =\mathbb{Z}_2$,}\\
\mathbb{Z}_4 & \textrm{if $N=4l+2$, $\Gamma =\mathbb{Z}_2$ or $\Gamma =\mathbb{Z}_4$,}\\
\mathbb{Z}_2\oplus\mathbb{Z}_2 & \textrm{if $N=4l$, \, $\Gamma =\mathbb{Z}_2$ \, or \, \, $\Gamma =\mathbb{Z}_2\oplus\mathbb{Z}_2$.}
\end{array} \right.
\end{equation}
The group $Sp(N)$ has the center $Z(Sp(N))=\mathbb{Z}_2$.

Among the exceptional groups only $E_6$ and $E_7$ are interesting for our purposes, the rest of them ($G_2$, $F_4$ and $E_8$) are simply-connected and have a trivial center. The real compact forms of $E_6$ and $E_7$ are not simply-connected. The universal cover of $E_6$ has the center $Z(\tilde{E}_6)=\mathbb{Z}_3$, and the universal cover of $E_7$ has the center $Z(\tilde{E}_7)=\mathbb{Z}_2$.

We consider the gauge group $G=\tilde{G}/\Gamma$. 
The center of the cover $Z(\tilde{G})\subset T$ is a subgroup of the maximal torus. $T$ is given by the elements $H=e^{ih} \in T$, where $h\in\mathfrak{h}$ is in the Cartan subalgebra.

The irreducible representations of $\tilde{G}$ are labeled by highest weight with $n$ independent elements, where $n$ is the rank of $\tilde{G}$: $(\mu_1, ..., \mu_{n})$. 

In most cases the center of $\tilde{G}$, or its proper subgroup $\Gamma$, is given by $\mathbb{Z}_m$ for some $m\in\mathbb{Z}$, and the calculation looks similar to the $\tilde{G} = SU(N)$ example. But in general $\Gamma$ can be represented by a product of $i$ cyclic groups:
$\Gamma = \mathbb{Z}_{m_1}\times......\times\mathbb{Z}_{m_i}$. 
We take the elements $\sum_i k_ic_i \in \mathfrak{h}$ in the Cartan subalgebra $\mathfrak{h}$ of $\tilde{G}$ and exponentiate them to get the central elements $C_{k_1....k_i} = e^{i\sum_i k_ic_i}\in Z(\tilde{G})$. Here we account for the structure of $\Gamma$: the index $i$ refers to the $i$-th factor in the product and the coefficient $k_i$ labels the elements inside each factor $\mathbb{Z}_{m_i}$. 

In terms of the basis of the Cartan subalgebra $h_j\in \mathfrak{h}$ we can express $\sum_i k_ic_i=\sum_i 2\pi \frac{k_i}{m_i}\sum_{j=1}^n (a_i)_j h_j$, where $n$ is the dimension of $\mathfrak{h}$, $m_i$ is the number of the elements in $\mathbb{Z}_{m_i}$ and $(a_i)_j$ are real linear coefficients describing $c_i$ and depending on the choice of a basis ${h_i}$. 

The representation $R_\mu$ of a central element $C_{k_1,...,k_i}$ is given by the formula: $R_{\mu}(C_{k_1,...,k_i}) = R_{\mu}(e^{i\sum_i k_ic_i}) = e^{i\sum_i k_iR_{\mu}(c_i)}$. 
The characters of the central elements in the representations $R_\mu$ are:

\begin{equation}\label{char}
\chi_{R_{\mu}}(C_{k_1,...,k_i}) = Tr (e^{i\sum_i k_iR_{\mu}(c_i)})= d_{R\mu}\cdot e^{i\sum_i k_i<\mu,c_i>}= d_{R\mu}\cdot e^{i 2\pi \sum_i \frac{k_i}{m_i}[\mu_i]}.
\end{equation}
Here we have rewritten the pairing $\sum_i k_i<\mu,c_i>$ in the following way:  $<\mu,\sum_i k_ic_i>= i2\pi \sum_i \frac{k_i}{m_i} \sum_{j=1}^n (a_i)_j\mu_j=i2\pi \sum_i \frac{k_i}{m_i} [\mu_i]$, where $(a_i)_j$ are linear coefficients producing different linear combinations of $\mu_j$s for each $k_i$-th element. The $i$ different linear combinations $\sum_{j=1}^n (a_i)_j\mu_j\in\mathbb{Z}$ define $i$ types of equivalence classes of the highest weight $\mu$: $[\sum_{j=1}^n (a_i)_j\mu_j]\equiv [\mu_i]\in \mathbb{Z}_{m_i}$, where $[\mu_i]$ is an integer modulo $m_i$.

Without Wilson surface the partition function for $G$ is given by:

\begin{equation}
Z_{YM}(\tau) = \sum_i\frac{1}{m_i}\sum_{k_i=0}^{m_i-1}\sum_{R_{\mu}(G)} d_{R_{\mu}}^{1-2g}e^{-\tau C_2(R_{\mu})}\chi_{R_{\mu}}(C_{k_1,...,k_i}),
 \end{equation}
 where the sum is over the representation $R_\mu$ of $G=\tilde{G}/\Gamma$ and $i$ coefficients $k_i$ label a central element in the subgroup $\Gamma$.
 
Using \eqref{char} we compute:  
 \begin{spacing}{1.55}
\begin{equation}\label{general}
\begin{array}{lll}
Z_{YM}(\tau) &=& \sum_i\frac{1}{m_i}\sum_{k_i=0}^{m_i-1} \sum_{R_{\mu}(G)} d_{R_{\mu}}^{2-2g}e^{-\tau C_2(R_{\mu})}\cdot e^{ik_i<\mu,c_i>}\\
 &=&\sum_{R_{\mu}(G)} d_{R_{\mu}}^{2-2g}e^{-\tau C_2(R_{\mu})} \sum_i\frac{1}{m_i}\sum_{k_i=0}^{m_i}e^{i2\pi \frac{k_i}{m_i}[\mu_i] }.
 \end{array}
 \end{equation}
 \end{spacing}
 Here each sum over $k_i$ in the second line is different from zero and is equal to $m_i$ only if $\mu_i =0$ mod $m_i$ (i.e. $[\mu_i]=0$) . 
 This condition corresponds to choosing only those representations of $\tilde{G}$ in which the elements $C_{k_1,....,k_i}\in\Gamma$ are all trivial, i.e. the representations of $G=\tilde{G}/\Gamma$: 
 \begin{equation}
 Z_{YM}(\tau) = \sum_{R_{\mu}(G=\tilde{G}/\Gamma)} d_{R_{\mu}}^{2-2g}e^{-\tau C_2(R_{\mu})}.
 \end{equation}

Now let us introduce a Wilson surface of weight $\lambda$. Just like any highest weight, $\lambda$ will belong to $i$ types of equivalence classes defined by the pairing $k_i<\lambda,c_i>=2\pi\frac{k_i}{m_i}\sum_{j=1}^n (a_i)_j \lambda_j$, where $(a_i)_j$ are real linear coefficients for the pairing with the $k_i$-th element and $\sum_{j=1}^n (a_i)_j \lambda_j$ is an integer. Then the weight $\lambda$ will be characterised by belonging to $i$ types of equivalence classes: $[\lambda_i] = [\sum_{j=1}^n (a_i)_j \lambda_j]\in \mathbb{Z}_{m_i}$. The partition function with a Wilson surface of weight $\lambda$ is given by:  
 \begin{spacing}{1.55}
\begin{equation}\label{gen}
\begin{array}{lll}
 Z^{\lambda}_{YM}(\tau) &=& \sum_i\frac{1}{m_i}\sum_{k_i=0}^{m_i-1}\sum_{R_{\mu}(G)} d_{R_{\mu}}^{1-2g}e^{-\tau C_2(R_\mu)}\chi_{R_{\mu}}(C_{k_1,...,k_i})\frac{\chi_{\lambda}(C_{k_1,...,k_i})}{d_\lambda} \\
 &=& \sum_i\frac{1}{m_i}\sum_{k_i=0}^{m_i-1}\sum_{R_{\mu}(G)} d_{R_{\mu}}^{2-2g}e^{-\tau C_2(R_\mu)}e^{ik_i<\mu,c_i>}   e^{ik_i<\lambda,c_i>} \\
 
 &=& \sum_i\frac{1}{m_i}\sum_{k_i=0}^{m_i-1}\sum_{R_{\mu}(G)} d_R^{2-2g}e^{-\tau C_2(R_\mu)}e^{i2\pi\frac{k_i}{m_i}[\mu_i] }e^{i2\pi\frac{k_i}{m_i}[\lambda_i]}\\
 &=& \sum_{R_{\mu}(G)} d_R^{2-2g}e^{-\tau C_2(R_\mu)} \sum_i\frac{1}{m_i}\sum_{k_i=0}^{m_i-1} e^{i2\pi \frac{k_i}{m_i}[\mu_i+\lambda_i]}.
 \end{array}
\end{equation}
\end{spacing}
Now each sum over $k_i$ is different from zero and is equal to $m_i$ only if $[\mu_i+\lambda_i] = 0$ for all i. This results in:

\begin{equation}\label{gen2}
Z^{\lambda}_{YM}(\tau)=\sum_{R_{\mu +\lambda}(G=\tilde{G}/\Gamma)} dim_{R_{\mu}}^{2-2g} e^{-\tau C_2(R_{\mu})},
\end{equation}
where the sum is over such representations $R_\mu(\tilde{G})$, that the representations of $\tilde{G}$ with the highest weight $\mu+\lambda$ would correspond to the representations of $G=\tilde{G}/\Gamma$. 

\subsection{Example of $G=Spin(4l)/(\mathbb{Z}_2\oplus\mathbb{Z}_2)$}

Now let us illustrate the formulas \eqref{gen}, \eqref{gen2} with an example of a gauge group with $\pi_1(G) \cong \Gamma = \mathbb{Z}_{m_1}\times......\times\mathbb{Z}_{m_i}$. The covering group $\tilde{G}=Spin(4l)$ has the center given by a product of two copies of $\mathbb{Z}_2$: $Z(Spin(4l)) = \mathbb{Z}_2\oplus\mathbb{Z}_2$. 
If we factorize by the entire center we get $G= Spin(4l)/\mathbb{Z}_2\oplus\mathbb{Z}_2$.  
We start with the central elements of $\tilde{G}= Spin(4l)$: $C_{k_1k_2} = e^{i(k_1c_1+k_2c_2)}\in \mathbb{Z}_2\oplus\mathbb{Z}_2$, where 
$k_1c_1+k_2c_2=\pi k_1\sum_{i=1}^n (a_1)_i h_i+\pi k_2\sum_{i=1}^n (a_2)_i h_i \in \mathfrak{h}$ are in the Cartan subalgebra of $Spin(4l)$, the coefficients $k_j=0,1$ label the elements in the j-th copy of $\mathbb{Z}_2$, and $(a_1)_i , (a_2)_i$ are real coefficients describing the elements $c_1$ and $c_2$ respectively and depending on the choice of a basis ${h_i}$. 

The representation $R_\mu$ of a central element $C_{k_1k_2}$ is given by the formula: $R_{\mu}(C_{k_1k_2}) = R_{\mu}(e^{i(k_1c_1+k_2c_2)}) = e^{i(k_1R_{\mu}(c_1)+k_2R_{\mu}(c_2))}$. 
The characters of the central elements in the representations $R_\mu$ are:
\begin{equation}
\chi_{R_{\mu}}(C_{k_1k_2}) = Tr(e^{i(k_1R_{\mu}(c_1)+k_2R_{\mu}(c_2))}) = d_{R\mu}\cdot e^{i(k_1<\mu,c_1>+k_2<\mu,c_2>)}=d_{R\mu}\cdot e^{i\pi (k_1[\mu_1]+k_2[\mu_2] )}.
\end{equation}

Here we\rq{}ve computed the pairing $<\mu,k_1c_1+k_2c_2>$ explicitly: $ i\pi (k_1\sum_{i=1}^n (a_1)_i\mu_i+k_2\sum_{i=1}^n (a_2)_i\mu_i)\equiv i\pi (k_1[\mu_1]+k_2[\mu_2] )$. We denote by $[\mu_1]$ and $[\mu_2]$ two different linear combinations ($\sum_{i=1}^n (a_1)_i\mu_i \in\mathbb{Z}$ and $\sum_{i=1}^n (a_2)_i\mu_i \in\mathbb{Z}$) of the components of the same highest weight $\mu$ modulo $2$.

Without Wilson surface the partition function for $G= Spin(4l)/\mathbb{Z}_2\oplus\mathbb{Z}_2$ is given by:
\begin{spacing}{1.55}
\begin{equation}\label{spin}
\begin{array}{lll}

 Z_{YM}(\tau) &=& \frac{1}{4}\sum_{k_1=0}^1 \sum_{k_2=0}^1\sum_{R_{\mu}(Spin(4l))} d_{R_{\mu}}^{1-2g}e^{-\tau C_2(R_{\mu})}\chi_{R_{\mu}}(C_{k_1k_2}) \\
 &=& \frac{1}{4}\sum_{k_1=0}^1 \sum_{k_2=0}^1 \sum_{R_{\mu}(Spin(4l))} d_{R_{\mu}}^{2-2g}e^{-\tau C_2(R_{\mu})}\cdot e^{i(k_1<\mu,c_1>+k_2<\mu,c_2>)}\\
 &=&\sum_{R_{\mu}(Spin(4l))} d_{R_{\mu}}^{2-2g}e^{-\tau C_2(R_{\mu})}\cdot \frac{1}{4}\sum_{k_1=0}^1\sum_{k_2=0}^1 e^{i\pi (k_1[\mu_1]+k_2[\mu_2] ) }\\
 &=& \sum_{R_{\mu}(Spin(4l)/(\mathbb{Z}_2\oplus\mathbb{Z}_2))} d_{R_{\mu}}^{2-2g}e^{-\tau C_2(R_{\mu})}.
 \end{array}
 \end{equation}
 \end{spacing}
In more detail, the sum in the first line of \eqref{spin} runs over the representations $R_\mu$ of $Spin(4l)$ and in the last line - over the representations $R_\mu$ of $Spin(4l)/(\mathbb{Z}_2\oplus\mathbb{Z}_2)$. The change happens for the following reason. Each sum over $k_i$ in the second line is equal to zero, or to $2$ if $\sum_{i=1}^n a_i\mu_i$ is even, i.e. $[\mu_i]=0$. This condition corresponds to chosing only those representations of $Spin(4l)$ in which the elements $C_{k_1k_2}\in \mathbb{Z}_2\oplus\mathbb{Z}_2$ are all trivial, i.e. the representations of $Spin(4l)/(\mathbb{Z}_2\oplus\mathbb{Z}_2)$.

When we introduce a Wilson surface of weight $\lambda$, it will involve defining two equivalence classes for $\lambda$ from the pairing  $<\lambda,k_ic_i>$: $[\lambda_1] = [\sum_{j=1}^n (a_1)_j \lambda_j]\in \mathbb{Z}_2$ and $[\lambda_2] = [\sum_{j=1}^n (a_2)_j\lambda_j]\in \mathbb{Z}_2$. The partition function in the presence of a Wilson surface is modified in the following way:  
 \begin{spacing}{1.55}
\begin{equation}
\begin{array}{lll}
 Z^{\lambda}_{YM}(\tau) &=& \frac{1}{4}\sum_{k_1=0}^1 \sum_{k_2=0}^1 \sum_{R_{\mu}(Spin(4l))} d_{R_{\mu}}^{1-2g}e^{-\tau C_2(R_\mu)}\chi_{R_{\mu}}(C_{k_1k_2})\frac{\chi_{\lambda}(C_{k_1k_2})}{d_\lambda} \\
 &=& \frac{1}{4}\sum_{k_1=0}^1 \sum_{k_2=0}^1 \sum_{R_{\mu}(Spin(4l))} d_{R_{\mu}}^{2-2g}e^{-\tau C_2(R-\mu)}e^{i<\mu,k_1c_1+k_2c_2>} e^{i<\lambda,k_1c_1+k_2c_2>} \\
 
 &=& \frac{1}{4}\sum_{k_1=0}^1 \sum_{k_2=0}^1 \sum_{R_{\mu}(Spin(4l))} d_R^{2-2g}e^{-\tau C_2(R)}e^{i(k_1[\mu_1]+k_2[\mu_2] )}e^{i(k_1[\lambda_1]+k_2[\lambda_2] )}\\
 &=& \sum_{R_{\mu}(Spin(4l))} d_R^{2-2g}e^{-\tau C_2(R)} \frac{1}{4}\sum_{k_1=0}^1 \sum_{k_2=0}^1 e^{i\pi (k_1[\mu_1+\lambda_1]+k_2 [\mu_2+\lambda_2])}\\
 &=& \sum_{\mu_1 \in[-\lambda_1], \, \mu_2 \in[-\lambda_2]} dim_{R_{\mu}}^{2-2g} e^{-\tau C_2(R_{\mu})},
 \end{array}
\end{equation}
\end{spacing}
Here the sum over each $k_i$ is different from zero only for $\mu_i+\lambda_i$ even. This condition reduces the sum in the last line to the sum over such representations $R_\mu(Spin(4l))$ that the representations of $Spin(4l)$ with the highest weight $\mu+\lambda$ would correspond to the representations of $Spin(4l)/(\mathbb{Z}_2\oplus\mathbb{Z}_2)$.


\end{document}